\documentclass[a4paper,10pt]{article}
\usepackage[dvips]{graphicx}
\usepackage{amsmath}
\usepackage{a4wide}

\newtheorem{condition}{Condition}

\newtheorem{corollary}{Corollary}

\newtheorem{example}{Example}

\newtheorem{lemma}{Lemma}

\newtheorem{proposition}{Proposition}
\newtheorem{remark}{Remark}

\newenvironment{proof}[1][Proof]{\textbf{#1.} }{\ \rule{0.5em}{0.5em}}
\makeatletter
\def\@removefromreset#1#2{\let\@tempb\@elt
\def\@tempa#1{@&#1}\expandafter\let\csname @*#1*\endcsname\@tempa
\def\@elt##1{\expandafter\ifx\csname @*##1*\endcsname\@tempa\else
\noexpand\@elt{##1}\fi}     \expandafter\edef\csname cl@#2\endcsname{\csname cl@#2\endcsname}     \let\@elt\@tempb
\expandafter\let\csname @*#1*\endcsname\@undefined}

\@removefromreset{equation}{section}

\@removefromreset{theorem}{section}
\makeatother
\input{tcilatex}

\begin{document}

\title{''Local Realism'', Bell's Theorem and \\
Quantum \textquotedblright Locally Realistic\textquotedblright\
Inequalities\bigskip }
\author{Elena R. Loubenets\bigskip \\
%EndAName
\textit{Department of Mathematical Sciences, University of }\\
\textit{Aarhus, Ny Munkegade, DK-8000, Aarhus, Denmark \smallskip }\\
\textit{e-mail:\ elena@imf.au.dk; erl@erl.msk.ru}\\
\bigskip \\
\bigskip}
\date{}
\maketitle

\baselineskip20pt

\bigskip

\noindent Based on the new general framework for the probabilistic
description of experiments, introduced in Ref. 6, 7, we analyze in
mathematical terms the link between the validity of Bell-type inequalities
under joint experiments upon a system of any type and the physical concept
of \textquotedblright local realism\textquotedblright . We prove that the
violation of Bell-type inequalities in the quantum case has no connection
with the violation of \textquotedblright local realism\textquotedblright .
In a general setting, we formulate in mathematical terms a condition on
\textquotedblright local realism\textquotedblright\ under a joint experiment
and consider examples of quantum \textquotedblright locally
realistic\textquotedblright\ joint experiments. We, in particular, show that
quantum joint experiments of the Alice/Bob type are \textquotedblright
locally realistic\textquotedblright . For an arbitrary bipartite quantum
state,\ we derive quantum analogs of the original Bell inequality. In view
of our results, we argue that the violation of Bell-type inequalities in the
quantum case cannot be a valid argument in the discussion on locality or
non-locality of quantum interactions.\newline
\newline

\section{INTRODUCTION}

\bigskip

The Bell inequality (Refs. 1, 2) and the\ Clauser-Horne-Shimony-Holt (%
\textit{CHSH}) inequality (Ref. 3) describe the relation between the
statistical data observed under joint measurements. The original derivations
of these inequalities (and their further numerous generalizations and
strengthenings) are based on the structure of probability theory\footnote{%
See, for example, in Ref. 4.} associated with the formalism of random
variables. The latter probabilistic formalism is often referred to as 
\textit{classical}\emph{\ }probability.

The sufficient mathematical condition used for the derivation of the above
inequalities in the \textit{classical} probabilistic frame is usually linked
with the physical concept of \textquotedblright \emph{local realism}%
\textquotedblright . The latter refers (see, for example, in Ref. 5, page
160) to those situations where, under a joint experiment, set-ups of
marginal experiments are chosen independently.

In the quantum case, the Bell inequality is, in general, violated and Bell's
theorem\footnote{%
As explained by A Shimony in private communication, the first published use
of the term \textit{Bell's theorem}\ appeared in Ref. 3 and referred to the
acknowledgment of J. Bell's results in Ref. 1.} states that a
\textquotedblright locally realistic\textquotedblright\ model cannot
describe statistics under joint quantum measurements.

In the present paper, we analyze this statement from the point of view of
the general framework for the probabilistic description of experiments
introduced in Refs. 6, 7. Based on the notions of an information state and a
generalized observable, this new probabilistic formalism allows to describe
both classical and quantum measurements in a unified way.

In Sec. 2, we discuss in a general setting the description of a joint
experiment performed on a system of any type represented initially by an
information state.

In Sec. 3, we formulate a general mathematical condition sufficient for the
validity a CHSH-form inequality under joint measurements\footnote{%
An experiment with real-valued outcomes is usually referred to as a
measurement.} upon a system of any type. This sufficient condition concerns
only a factorizable form of joint generalized observables describing the
corresponding joint measurements and does not, in general, result in the
existence of a local hidden variable model for a system information state.
We underline that though joint generalized observables describing classical
joint measurements are factorizable, the converse is not true and
factorizable generalized observables may represent quantum joint
measurements.

For factorizable generalized observables, we further specify the general
condition sufficient for the validity of the original Bell inequality%
\footnote{%
Our derivation of the Bell inequality is valid for any type of outcomes and
does not exploit generally accepted \textquotedblright measurement
result\textquotedblright\ restrictions introduced in Ref. 1, 2.}. We prove
that Bell's correlation restriction on the observed outcomes (see in Refs.
1, 2) represents only a particular case of the general sufficient condition
that we introduce in this paper. Under the latter sufficient condition, the
Bell inequality holds even if the observed outcomes are not perfectly
correlated or anticorrelated.

We discuss possible mathematical reasons for the violation of a CHSH-form
inequality and point out that the sufficient condition for its validity does
not, in general, represent mathematically the physical concept of
\textquotedblright local realism\textquotedblright .

In Sec. 4, we formulate in mathematical terms a general condition on
\textquotedblright local realism\textquotedblright\ under a joint experiment
upon a system of any type and consider examples of quantum
\textquotedblright locally realistic\textquotedblright \footnote{%
If, under a joint experiment, the physical concept of "local realism" is not
violated then, for short, we refer to this joint experiment as "locally
realistic".}\ joint experiments. We, in particular, show that quantum joint
measurements of the Alice/Bob type are "locally realistic\textquotedblright .

From our presentation it follows that, under \textquotedblright locally
realistic\textquotedblright\ joint measurements, a CHSH-form inequality may
be violated whenever joint generalized observables, describing these
measurements, do not have a factorizable form. The latter is just a general
situation under quantum joint measurements of Alice and Bob.

In Sec. 5, for an arbitrary bipartite quantum state, we derive quantum
analogs of the original Bell inequality.

In Sec. 6, we argue that the violation of Bell-type inequalities under
quantum Alice/Bob joint measurements does not point to non-locality of
quantum interactions.

\bigskip

\bigskip

\section{DESCRIPTION\ OF JOINT\ MEASUREMENTS}

\bigskip

Consider the description of an experiment with outcomes in a set $\Lambda $
and performed on a system of any type.

Let, before an experiment, a system be characterized in terms of some
properties $\theta \in \Theta $ of any nature and the uncertainty of
possible $\theta $ be specified by a $\sigma $-algebra $\mathcal{F}_{\Theta
} $ of subsets of $\Theta $ and a probability distribution $\pi $ on $%
(\Theta ,\mathcal{F}_{\Theta })$.

We refer to a measurable space $(\Theta ,\mathcal{F}_{\Theta })$ as a system 
\emph{information space} and call a triple 
\begin{equation}
\mathcal{I}:=(\Theta ,\mathcal{F}_{\Theta },\pi )  \label{1}
\end{equation}%
an \emph{information state} of a system (see Refs. 6, 7, for details). We
say that an information state $\mathcal{I}$ has the support on a set $F\in 
\mathcal{F}_{\Theta }$ if $\pi (F)=1$.

The above mathematical setting on initial representation of a system is
rather general and covers a broad class of probabilistic situations arising
under the description of experiments, in particular, those of \textit{%
classical} probability, of quantum measurement theory and, more generally,
all those situations where each $\theta $ is interpreted as a
\textquotedblright bit\textquotedblright\ of information available on a
system and the uncertainty of possible \textquotedblright
bits\textquotedblright\ is specified by a probability distribution $\pi $.

According to our consideration in Ref. 6, 7, any experiment, with an outcome
set $\Lambda $, performed on a system, described initially by an information
space $(\Theta ,\mathcal{F}_{\Theta })$, is uniquely represented on this
information space by a \emph{generalized observable }$\Pi $. If $\mathcal{I}$
is a system initial information state then the probability $\mu ^{(\Pi )}(D;%
\mathcal{I})$ that an outcome $\lambda $ belongs to a subset $B$ of $\Lambda 
$ is given by (see Ref. 6, 7): 
\begin{equation}
\mu ^{(\Pi )}(B;\mathcal{I})=\int_{\Theta }(\Pi (B))(\theta )\pi (d\theta ),
\label{2}
\end{equation}
where: (i) for any outcome subset $B\subseteq \Lambda $, the real-valued
function $(\Pi (B))(\cdot )$ on $(\Theta ,\mathcal{F}_{\Theta })$ is
measurable, with values in $[0,1];$ (ii) for any $\theta \in \Theta ,$ the
mapping $(\Pi (\cdot ))(\theta )$ represents a probability distribution of
outcomes in $\Lambda .$ Thus, $\Pi $ is a normalized measure with values $%
\Pi (B),$ $\forall B\subseteq \Lambda ,$ that are nonnegative real-valued
measurable functions on $(\Theta ,\mathcal{F}_{\Theta })$. By its measure
structure, $\Pi $ is similar to the notion of a positive operator-valued (%
\textit{POV}) measure in quantum measurement theory\footnote{%
For quantum measurement theory, see Refs. 8 - 10.}.

If a system initial information space $(\Theta ,\mathcal{F}_{\Theta })$
provides \textquotedblright \emph{no knowledge\textquotedblright }\footnote{%
In this case, the probability distribution of outcomes does not depend on
information on a system specified by $(\Theta ,\mathcal{F}_{\Theta }).$} on
an experiment upon this system then this experiment is represented on $%
(\Theta ,\mathcal{F}_{\Theta })$ by a \textit{trivial }generalized
observable $(\Pi (B))(\theta )=\mu ^{\Pi }(B),$ $\forall \theta \in \Theta .$

Let $\Pi ,$ with an outcome set $\Lambda _{1}\times \Lambda _{2},$ be a
generalized observable, representing on $(\Theta ,\mathcal{F}_{\Theta })$ a
joint\footnote{%
An experiment with outcomes in a product set $\Lambda _{1}\times \Lambda
_{2} $ is called \emph{joint}.} experiment. For any $B_{1}\subseteq \Lambda
_{1}$, $B_{2}\subseteq \Lambda _{2},$ the relations 
\begin{equation}
\Pi _{1}(B_{1}):=\Pi (B_{1}\times \Lambda _{2}),\text{ \ \ \ \ \ }\Pi
_{2}(B_{2}):=\Pi (\Lambda _{1}\times B_{2})  \label{3}
\end{equation}
define the generalized observables $\Pi _{1},$ with the outcome set $\Lambda
_{1},$ and $\Pi _{2}$, with the outcome set $\Lambda _{2}.$ Each of the
latter generalized observables is called \emph{marginal} and represents on $%
(\Theta ,\mathcal{F}_{\Theta })$ the corresponding \emph{marginal }experiment%
\footnote{%
That is,\emph{\ }such an experimental situation where under a joint
experiment outcomes either in $\Lambda _{2}$ or in $\Lambda _{1}$ are
ignored completely.}. With respect to $\Pi _{1}$ and $\Pi _{2}$, the
generalized observable $\Pi $ is called \emph{joint.}

We further consider joint experiments with real-valued outcomes $\lambda
_{i}\in \Lambda _{i}$ (that is, joint measurements) and, for simplicity,
suppose that outcomes are bounded $|\lambda _{i}|\leq \mathrm{C}_{i},$ $%
i=1,2 $.

Under a joint measurement performed on a system in an initial information
state $\mathcal{I},$ consider the expectation values 
\begin{equation}
\langle \lambda _{i}\rangle _{_{\mathcal{I}}}^{(\Pi )}:=\int_{\Lambda
_{1}\times \Lambda _{2}}\lambda _{i}\mu ^{(\Pi )}(d\lambda _{1}\times
d\lambda _{2};\mathcal{I}),\text{ \ \ }i=1,2,  \label{4}
\end{equation}%
of the observed outcomes $\lambda _{i}\in \Lambda _{i}$ and the expectation
value 
\begin{equation}
\langle \lambda _{1}\lambda _{2}\rangle _{_{\mathcal{I}}}^{(\Pi
)}:=\int_{\Lambda _{1}\times \Lambda _{2}}\lambda _{1}\lambda _{2}\mu ^{(\Pi
)}(d\lambda _{1}\times d\lambda _{2};\mathcal{I})  \label{5}
\end{equation}%
of the product $\lambda _{1}\lambda _{2}$ of the observed outcomes. Due to
Eqs. (\ref{2}) and (\ref{3}), we have: 
\begin{align}
\langle \lambda _{i}\rangle _{_{\mathcal{I}}}^{(\Pi )}& =\int_{\Lambda
_{i}}\int_{\Theta }\lambda _{i}(\Pi _{i}(d\lambda _{i}))(\theta )\pi
(d\theta )  \label{6} \\
& =\int_{\Theta }f_{i}(\theta )\pi (d\theta ),\text{ \ \ \ \ \ }\ \ i=1,2, 
\notag \\
&  \notag \\
\langle \lambda _{1}\lambda _{2}\rangle _{_{\mathcal{I}}}^{(\Pi )}&
=\int_{\Lambda _{1}\times \Lambda _{2}}\int_{\Theta }\lambda _{1}\lambda
_{2}(\Pi (d\lambda _{1}\times d\lambda _{2}))(\theta )\pi (d\theta )
\label{7} \\
& =\int_{\Theta }f_{joint}(\theta )\pi (d\theta ),  \notag
\end{align}%
where 
\begin{align}
f_{i}(\theta )& :=\int_{\Lambda _{i}}\lambda _{i}(\Pi _{i}(d\lambda
_{i}))(\theta ),\text{ \ \ \ \ \ \ \ \ \ \ \ \ \ \ \ \ \ \ \ \ \ \ \ }%
|f_{i}(\theta )|\leq \mathrm{C}_{i},\text{ \ \ }\forall \theta \in \Theta ,%
\text{ \ \ }i=1,2,  \label{8} \\
f_{joint}(\theta )& :=\int_{\Lambda _{1}\times \Lambda _{2}}\lambda
_{1}\lambda _{2}(\Pi (d\lambda _{1}\times d\lambda _{2}))(\theta ),\text{ \
\ \ }|f_{joint}(\theta )|\leq \mathrm{C}_{1}\mathrm{C}_{2},\text{ \ \ }%
\forall \theta \in \Theta ,  \label{9}
\end{align}%
are real-valued measurable functions on $(\Theta ,\mathcal{F}_{\Theta }).$
(In probability theory, a measurable real-valued function is usually
referred to as a random variable, see, for example, in Ref. 4.)

From Eqs. (\ref{6}) - (\ref{9}) it follows that, under a joint measurement
upon a system of any type represented initially by an information state $%
\mathcal{I},$ the expectation values are always expressed in terms of random
variables on $(\Theta ,\mathcal{F}_{\Theta }).$

However, in contrast to the generally accepted formalism of \textit{%
probability theory }(see, for example, in Ref. 4), the values of random
variables $f_{i}$, $i=1,2,$ in Eq. (\ref{6}) do not, in general, represent
outcomes observed under this joint measurement while the random variable $%
f_{joint}$ in Eq. (\ref{7}) does not, in general, coincide with the product $%
f_{1}f_{2}$.

\begin{remark}[On classical measurements]
The values of the random variables $f_{1}$ and $f_{2}$ in (\ref{6}) do
represent the observed real-valued outcomes iff\emph{\ }a joint generalized
observable has the \textquotedblright image\textquotedblright\ form (see in
Ref. 6, 7), that is: 
\begin{eqnarray}
(\Pi ^{(image)}(B_{1}\times B_{2}))(\theta ) &=&\chi _{f_{1}^{-1}(B_{1})\cap
f_{2}^{-1}(B_{2})}(\theta )  \label{10} \\
&=&\chi _{f_{1}^{-1}(B_{1})}(\theta )\chi _{f_{2}^{-1}(B_{2})}(\theta ), 
\notag
\end{eqnarray}%
for any $\theta \in \Theta $ and any outcome subsets $B_{1}\subseteq \Lambda
_{1},$ $B_{2}\subseteq \Lambda _{2}.$ Here, $\chi _{F}(\theta )$ is an
indicator function\footnote{%
That is: $\chi _{F}(\theta )=1,$ $\forall \theta \in F$ and $\chi
_{F}(\theta )=0,$ $\forall \theta \notin F.$} of a set $F\in \mathcal{F}%
_{\Theta }$ and 
\begin{equation}
f_{i}^{-1}(B_{i}):=\{\theta \in \Theta :f_{i}(\theta )\in B_{i}\}  \label{11}
\end{equation}%
is the preimage in $\mathcal{F}_{\Theta }$ of a subset $B_{i}\subseteq
\Lambda _{i}$. \textquotedblright Image\textquotedblright\ generalized
observables describe ideal measurements on a classical system (usually
referred to as classical measurements).\newline
Due to Eqs. (\ref{2}), (\ref{7}) - (\ref{10}), under a classical joint
measurement, the probability distribution $\mu ^{(cl)}$ of outcomes in $%
\Lambda _{1}\times \Lambda _{2}$ has the \textquotedblright
image\textquotedblright\ form: 
\begin{equation}
\mu ^{(cl)}(B_{1}\times B_{2};\mathcal{I})=\pi (f_{1}^{-1}(B_{1})\cap
f_{2}^{-1}(B_{2})),  \label{12}
\end{equation}%
the random variable $f_{joint}=$ $f_{1}f_{2}$ and the expectation value of
the product of outcomes is given by: 
\begin{equation}
\langle \lambda _{1}\lambda _{2}\rangle _{_{\mathcal{I}}}^{(clas)}=\int_{%
\Theta }f_{1}(\theta )f_{2}(\theta )\pi (d\theta ).  \label{13}
\end{equation}
\end{remark}

\medskip

As we discuss this in detail in Sec. 3.1, the relation $f_{joint}=$ $%
f_{1}f_{2}$ holds not only for a classical joint measurement but for any
joint measurement described on $(\Theta ,\mathcal{F}_{\Theta })$ by a
generalized observable of the product form (see Eq. (\ref{19})). However, in
the latter case, the values of$\ f_{1}$ and $f_{2}$ do not, in general,
represent the observed outcomes.

Consider now two joint measurements, performed on a system of any type and
represented on a system information space $(\Theta ,\mathcal{F}_{\Theta })$
by joint generalized observables $\Pi ^{(1)}$ and $\Pi ^{(2)}.$ From Eq. (%
\ref{7}) it follows: 
\begin{equation}
\langle \lambda _{1}\lambda _{2}\rangle _{_{\mathcal{I}}}^{^{(\Pi
^{(1)})}}\pm \langle \lambda _{1}\lambda _{2}\rangle _{_{\mathcal{I}%
}}^{^{(\Pi ^{(2)})}}=\int_{\Theta }\{f_{joint}^{(1)}(\theta )\pm
f_{joint}^{(2)}(\theta )\}\pi (d\theta ).  \label{14}
\end{equation}%
Due to the relation 
\begin{equation}
\left\vert x-y\right\vert \leq 1-xy,  \label{15}
\end{equation}%
valid for any real numbers $|x|\leq 1,$ $|y|\leq 1,$ the inequality 
\begin{equation}
\left\vert \langle \lambda _{1}\lambda _{2}\rangle _{_{\mathcal{I}}}^{^{(\Pi
^{(1)})}}\pm \langle \lambda _{1}\lambda _{2}\rangle _{_{\mathcal{I}%
}}^{^{(\Pi ^{(2)})}}\right\vert \leq \mathrm{C}_{1}\mathrm{C}_{2}\pm \frac{1%
}{\mathrm{C}_{1}\mathrm{C}_{2}}\langle \lambda _{1}\lambda _{2}\rangle _{_{%
\mathcal{I}}}^{^{(\Pi ^{(1)})}}\langle \lambda _{1}\lambda _{2}\rangle _{_{%
\mathcal{I}}}^{^{(\Pi ^{(2)})}}  \label{16}
\end{equation}%
holds for any information state $\mathcal{I}$ and any generalized
observables $\Pi ^{(1)}$ and $\Pi ^{(2)}.$

In the following section, we derive an upper bound of the expression (\ref%
{14}) for joint generalized observables of the special type.

\bigskip

\bigskip

\section{FACTORIZABLE GENERALIZED OBSERVABLES}

\bigskip

We say that a joint generalized observable $\Pi $ on an information space $%
(\Theta ,\mathcal{F}_{\Theta })$ is \emph{factorizable on a set }$F\in 
\mathcal{F}_{\Theta }$ if $\Pi $ admits a representation 
\begin{equation}
(\Pi (B_{1}\times B_{2}))(\theta )=\int_{\Omega }(\Pi _{1,\omega
}(B_{1}))(\theta )(\Pi _{2,\omega }(B_{2}))(\theta )\text{ }\nu (d\omega ),
\label{17}
\end{equation}%
for any $\theta \in F$ and any outcome subsets $B_{i}$ $\subseteq \Lambda
_{i},$ $i=1,2.$ Here: (i) $(\Omega ,\mathcal{F}_{\Omega })$ is some
measurable space; (ii) $\nu $ is a probability distribution on $(\Omega ,%
\mathcal{F}_{\Omega })$; (iii) $\Pi _{1,\omega }$ and $\Pi _{2,\omega }$ are
generalized observables on $(\Theta ,\mathcal{F}_{\Theta })$ with outcome
sets $\Lambda _{1}$ and $\Lambda _{2},$ respectively. To express Eq. (\ref%
{17}) in short, we use the notation\footnote{%
In measure theory, the notation $\mu \times \nu $ is generally accepted for
the product measure with the marginal measures $\mu $ and $\nu .$} 
\begin{equation}
\Pi \overset{F}{=}\int_{\Omega }\Pi _{1,\omega }\times \Pi _{2,\omega }\text{
}\nu (d\omega ),  \label{18}
\end{equation}%
and we omit "$F$" whenever $\Pi $ is factorizable on all of a set $\Theta $.

If, in particular, $\nu =\delta _{\omega _{0}},$ $\forall \omega _{0}\in
\Omega ,$ is a Dirac measure, then, in Eq. (\ref{18}), a joint generalized
observable 
\begin{equation}
\Pi \overset{F}{=}\Pi _{1,\omega _{0}}\times \Pi _{2,\omega _{0}}  \label{19}
\end{equation}
has the product form on a set $F\in \mathcal{F}_{\Theta }$, with the
generalized observables $\Pi _{1,\omega _{0}}$ and $\Pi _{2,\omega _{0}}$
representing the marginal experiments.

Notice that an "\textit{image" }generalized observable (\ref{10}),
representing on $(\Theta ,\mathcal{F}_{\Theta })$ a classical joint
measurement, is of the product form on all of $\Theta $.

\subsection{Bell-type Inequalities}

\bigskip

For simplicity, we first consider the case of product generalized
observables.

For a joint measurement with outcomes $\left\vert \lambda _{1}\right\vert
\leq \mathrm{C}_{1}$, $\left\vert \lambda _{2}\right\vert \leq \mathrm{C}%
_{2} $, let the corresponding generalized observable on $(\Theta ,\mathcal{F}%
_{\Theta })$ be product (see Eq. (\ref{19})) and have the form: 
\begin{equation}
\Pi ^{(a,b)}\overset{F}{=}\Pi _{1}^{(a)}\times \Pi _{2}^{(b)},\text{ \ }\ \ 
\text{\ }\forall F\in \mathcal{F}_{\Theta }.  \label{20}
\end{equation}
Here, in the left-hand side, a parameter standing in the first place of a
pair specifies a set-up of the marginal measurement with outcomes in $%
\Lambda _{1}$ while a parameter standing in the second place of a pair - a
set-up of the marginal measurement with outcomes in $\Lambda _{2}.$ In the
right-hand, the lower indices refer to outcome sets $\Lambda _{1}$ and $%
\Lambda _{2}$.

For a generalized observable of the form (\ref{20}), the corresponding
random variable $f_{joint}^{(a,b)}$ in Eq. (\ref{9}) has the product form: 
\begin{equation}
f_{joint}^{(a,b)}(\theta )=f_{1}(\theta ,a)f_{2}(\theta ,b),\text{ \ \ \ }%
\forall \theta \in F,  \label{21}
\end{equation}%
where 
\begin{equation}
f_{1}(\theta ,a)=\int_{\Lambda _{1}}\lambda _{1}(\Pi _{1}^{(a)}(d\lambda
_{1}))(\theta ),\text{ \ \ \ }f_{2}(\theta ,b)=\int_{\Lambda _{2}}\lambda
_{2}(\Pi _{2}^{(b)}(d\lambda _{2}))(\theta )  \label{22}
\end{equation}%
and $|f_{1}(\theta ,a)|\leq \mathrm{C}_{1}\mathrm{,}\ |f_{2}(\theta ,b)|\leq 
\mathrm{C}_{2},$ for any $\theta \in F.$

From Eq. (\ref{21}) it follows that, under a joint measurement, represented
on a system information space $(\Theta ,\mathcal{F}_{\Theta })$ by a product
generalized observable of the form (\ref{20}), the expectation value (\ref{7}%
) admits the representation: 
\begin{equation}
\langle \lambda _{1}\lambda _{2}\rangle _{_{\mathcal{I}}}^{(a,b)}:=\langle
\lambda _{1}\lambda _{2}\rangle _{\mathcal{I}}^{^{(\Pi
^{(a.b)})}}=\int_{F}f_{1}(\theta ,a)f_{2}(\theta ,b)\pi (d\theta ),
\label{23}
\end{equation}
for any initial information state $\mathcal{I}$ with the support on $F\in 
\mathcal{F}_{\Theta }.$

\begin{lemma}
For two product generalized observables of the form (\ref{20}), the
corresponding expectation values satisfy the relation 
\begin{equation}
\left\vert \gamma _{1}\langle \lambda _{1}\lambda _{2}\rangle _{_{\mathcal{I}%
}}^{(a,b_{1})}+\gamma _{2}\langle \lambda _{1}\lambda _{2}\rangle _{_{%
\mathcal{I}}}^{(a,b_{2})}\right\vert \leq \mathrm{C}_{1}\mathrm{C}%
_{2}+\gamma _{1}\gamma _{2}\frac{\mathrm{C}_{1}}{\mathrm{C}_{2}}\langle
\lambda _{2}^{\prime }\lambda _{2}\rangle _{_{\mathcal{I}}}^{^{(\widetilde{%
\Pi })}},  \label{24}
\end{equation}%
for any information state $\mathcal{I}$ with the support on $F\in \mathcal{F}%
_{\Theta }$ and any real-valued coefficients $\left\vert \gamma
_{1}\right\vert \leq 1,$ $\left\vert \gamma _{2}\right\vert \leq 1.$ Here, 
\begin{align}
\langle \lambda _{2}^{\prime }\lambda _{2}\rangle _{_{\mathcal{I}}}^{^{(%
\widetilde{\Pi })}}& :=\int_{F}f_{2}(\theta ,b_{1})f_{2}(\theta ,b_{2})\pi
(d\theta )  \label{25} \\
& =\int_{\Lambda _{2}\times \Lambda _{2}}\int_{F}\lambda _{2}^{\prime
}\lambda _{2}(\widetilde{\Pi }(d\lambda _{2}^{\prime }\times d\lambda
_{2}))(\theta )\pi (d\theta )  \notag
\end{align}%
and $\widetilde{\Pi }:\overset{F}{=}\Pi _{2}^{(b_{1})}\times \Pi
_{2}^{(b_{2})}.$
\end{lemma}

\begin{proof}
For a state $\mathcal{I}$ with the support on $F\in \mathcal{F}_{\Theta },$
the proof is based on the representation (\ref{23}), the inequality (\ref{15}%
), the relation $\pi (F)=1,$ Eq. (\ref{22}) and the notation (\ref{25}).
Specifically: 
\begin{align}
\left\vert \gamma _{1}\langle \lambda _{1}\lambda _{2}\rangle _{_{\mathcal{I}%
}}^{(a,b_{1})}+\gamma _{2}\langle \lambda _{1}\lambda _{2}\rangle _{_{%
\mathcal{I}}}^{(a,b_{2})}\right\vert & \leq \int_{F}\left\vert f_{1}(\theta
,a)\{\gamma _{1}f_{2}(\theta ,b_{1})+\gamma _{2}f_{2}(\theta
,b_{2})\}\right\vert \pi (d\theta )  \label{26} \\
& \leq \mathrm{C}_{1}\int_{F}\left\vert \gamma _{1}f_{2}(\theta
,b_{1})+\gamma _{2}f_{2}(\theta ,b_{2})\right\vert \pi (d\theta )  \notag \\
& \leq \mathrm{C}_{1}\mathrm{C}_{2}\int_{F}\{1+\frac{\gamma _{1}\gamma _{2}}{%
\mathrm{C}_{2}^{2}}f_{2}(\theta ,b_{1})f_{2}(\theta ,b_{2})\}\pi (d\theta ) 
\notag \\
& =\mathrm{C}_{1}\mathrm{C}_{2}+\gamma _{1}\gamma _{2}\frac{\mathrm{C}_{1}}{%
\mathrm{C}_{2}}\langle \lambda _{2}^{\prime }\lambda _{2}\rangle _{_{%
\mathcal{I}}}^{^{(\widetilde{\Pi })}}.  \notag
\end{align}
\end{proof}

\smallskip

From Lemma 1 it follows the following general statement. (For simplicity, we
further consider the case $\mathrm{C}_{1}=\mathrm{C}_{2}=1).$

\begin{proposition}[The Bell inequality]
Let a system be represented initially by an information state $\mathcal{I}%
=(\Theta ,\mathcal{F}_{\Theta },\pi )$ with the support on $F\in \mathcal{F}%
_{\Theta }$ and three joint measurements, with outcomes $\left\vert \lambda
_{1}\right\vert \leq 1$ and $\left\vert \lambda _{2}\right\vert \leq 1,$
performed on this system, be described on $(\Theta ,\mathcal{F}_{\Theta })$
by the product generalized observables of the form (\ref{20}), specified by
pairs $(a,b_{1}),$ $(a,b_{2})$ and $(b_{1},b_{2})$ of measurement
parameters. If 
\begin{eqnarray}
\int_{\Lambda _{2}}\lambda _{2}(\mathrm{\Pi }_{2}^{(b_{1})}(d\lambda
_{2}))(\theta ) &=&\pm \int_{\Lambda _{1}}\lambda _{1}(\mathrm{\Pi }%
_{1}^{(b_{1})}(d\lambda _{1}))(\theta )\text{ \ }\Longleftrightarrow \text{
\ \ \ }  \label{27} \\
f_{2}(\theta ,b_{1}) &=&\pm f_{1}(\theta ,b_{1}),  \notag
\end{eqnarray}%
$\pi $-almost everywhere\footnote{%
The term \textquotedblright $\pi $-almost everywhere on $F$%
\textquotedblright\ (a.e., for short) means that some relation holds on $F$
excluding the null sets of a probability distribution $\pi $.} on $F,$ then 
\begin{equation}
\left\vert \langle \lambda _{1}\lambda _{2}\rangle _{_{\mathcal{I}%
}}^{(a,b)}-\langle \lambda _{1}\lambda _{2}\rangle _{_{\mathcal{I}%
}}^{(a,b_{2})}\right\vert \leq 1\mp \langle \lambda _{1}\lambda _{2}\rangle
_{_{\mathcal{I}}}^{(b_{1},b_{2})}.  \label{28}
\end{equation}
\end{proposition}

\bigskip

Notice that, in Proposition 1, we derive the original Bell inequality (\ref%
{28}) without the so-called \textquotedblright measurement
result\textquotedblright\ restrictions on the observed outcomes, introduced
in Refs. 1, 2 and generally accepted in the literature on the Bell
inequality.

\begin{proposition}
Let four joint measurements, with outcomes $\left| \lambda _{1}\right| \leq
1 $ and $\left| \lambda _{2}\right| \leq 1$, be represented by the product
generalized observables of the form (\ref{20}), specified by pairs $%
(a_{k},b_{m}),$ $\forall k,m=1,2,$ of measurement parameters. Then the
extended CHSH inequality\footnote{%
Introduced in Ref. 12.} 
\begin{equation}
|\sum_{k,m=1,2}\gamma _{km}\langle \lambda _{1}\lambda _{2}\rangle _{_{%
\mathcal{I}}}^{(a_{k},b_{m})}|\leq 2  \label{29}
\end{equation}
holds for any initial information state $\mathcal{I}$ with the support on $%
F\in \mathcal{F}_{\Theta }$ and any real-valued coefficients $\left| \gamma
_{km}\right| \leq 1,$ $\forall k,m=1,2,$ such that $\gamma _{11}\gamma
_{12}=-\gamma _{21}\gamma _{22}$ or $\gamma _{11}\gamma _{21}=-\gamma
_{12}\gamma _{22}.$
\end{proposition}

\begin{proof}
In view of Lemma 1, we have: 
\begin{eqnarray}
|\sum_{k,m=1,2}\gamma _{km}\langle \lambda _{1}\lambda _{2}\rangle _{_{%
\mathcal{I}}}^{(a_{k},b_{m})}\text{ }|\text{ } &\leq &|\sum_{m=1,2}\gamma
_{1m}\langle \lambda _{1}\lambda _{2}\rangle _{_{\mathcal{I}%
}}^{(a_{1},b_{m})}\text{ }|+|\sum_{m=1,2}\gamma _{2m}\langle \lambda
_{1}\lambda _{2}\rangle _{_{\mathcal{I}}}^{(a_{2},b_{m})}|  \label{30} \\
&\leq &2+(\gamma _{11}\gamma _{12}+\gamma _{21}\gamma _{22})\langle \lambda
_{2}^{\prime }\lambda _{2}\rangle _{_{\mathcal{I}}}^{^{(\widetilde{\Pi })}}.
\notag
\end{eqnarray}%
The latter relation proves the statement for the case $\gamma _{11}\gamma
_{12}=-\gamma _{21}\gamma _{22}.$ Combining in the left hand side of the
inequality (\ref{29}) the first term with the third and the second term with
the fourth, we prove, quite similarly, the statement for the case $\gamma
_{11}\gamma _{21}=-\gamma _{12}\gamma _{22}.$
\end{proof}

\medskip

Clearly, the extended CHSH inequality is always true under classical joint
measurements\footnote{%
"Image" generalized observable describing classical joint measurements are
product, see Eq. (\ref{10}).} and the original derivation of the CHSH
inequality in Ref. 3 just corresponds to the classical case.

Consider now a more general situation where four joint measurements,
specified by pairs $(a_{k},b_{m}),$ $\forall k,m=1,2,$ of measurement
parameters, are represented on $(\Theta ,\mathcal{F}_{\Theta })$ by \emph{%
factorizable} generalized observables (see Eq. (\ref{18})) of the form: 
\begin{equation}
\Pi ^{(a_{k},b_{m})}\overset{F}{=}\int_{\Omega }\Pi _{1,\omega
}^{(a_{k})}\times \Pi _{2,\omega }^{(b_{m})}\text{ }\nu
_{a_{1},a_{2}}^{(b_{1},b_{2})}(d\omega ),\text{ \ \ }\ \forall k,m=1,2,\ 
\text{\ \ }\forall F\in \mathcal{F}_{\Theta },  \label{31}
\end{equation}%
where, in general, a probability distribution $\nu
_{a_{1},a_{2}}^{(b_{1},b_{2})}$ depends on set-ups of marginal measurements.
Under these joint measurements, the expectation values (\ref{7}) admit the
representations: 
\begin{eqnarray}
\langle \lambda _{1}\lambda _{2}\rangle _{_{\mathcal{I}}}^{(a_{k},b_{m})}
&=&\int_{F}\int_{\Omega }f_{1}(\theta ,\omega ,a_{k})f_{2}(\theta ,\omega
,b_{m})\nu _{a_{1},a_{2}}^{(b_{1},b_{2})}(d\omega )\pi (d\theta ),\text{ \ \
\ }\forall k,m=1,2,  \label{32} \\
f_{1}(\theta ,\omega ,a_{k}) &=&\int_{\Lambda _{1}}\lambda _{1}(\Pi
_{1,\omega }^{(a_{k})}(d\lambda _{1}))(\theta ),\text{ \ \ \ }f_{2}(\theta
,\omega ,b_{m})=\int_{\Lambda _{2}}\lambda _{2}(\Pi _{2,\omega
}^{(b_{m})}(d\lambda _{2}))(\theta ),  \notag
\end{eqnarray}%
for any information state $\mathcal{I}$ with the support on $F\in \mathcal{F}%
_{\Theta }.$

By its structure, these representations are quite similar to the
representation (\ref{23}). That is why, the above propositions can be easily
generalized.

\begin{proposition}[The extended CHSH inequality]
Under four joint measurements, described by factorizable generalized
observables (\ref{31}), the corresponding expectation values satisfy the
extended CHSH inequality (\ref{29}) for any initial information state $%
\mathcal{I}$ with the support on $F\in \mathcal{F}_{\Theta }.$
\end{proposition}

\medskip

Furthermore, let three joint measurements be described by factorizable
generalized observables of the form: 
\begin{eqnarray}
\Pi ^{(a,b_{m})}\overset{F}{=}\int_{\Omega }\Pi _{1,\omega }^{(a)}\times \Pi
_{2,\omega }^{(b_{m})}\text{ }\nu _{a}^{(b_{1},b_{2})}(d\omega ),\text{ \ \ }%
\ \forall F &\in &\mathcal{F}_{\Theta },\text{ \ \ }\forall m=1,2,
\label{33} \\
\Pi ^{(b_{1},b_{2})}\overset{F}{=}\int_{\Omega }\Pi _{1,\omega
}^{(b_{1})}\times \Pi _{2,\omega }^{(b_{2})}\text{ }\nu
_{a}^{(b_{1},b_{2})}(d\omega ),\text{ \ \ }\ \forall F &\in &\mathcal{F}%
_{\Theta }.  \notag
\end{eqnarray}%
It is easy to prove\footnote{%
Quite similarly to our proof of Proposition 1.} that, for these three joint
measurements, the corresponding expectation values 
\begin{eqnarray}
\langle \lambda _{1}\lambda _{2}\rangle _{_{\mathcal{I}}}^{(a,b_{m})}
&=&\int_{F}\int_{\Omega }f_{1}(\theta ,\omega ,a)f_{2}(\theta ,\omega
,b_{m})\nu _{a}^{(b_{1},b_{2})}(d\omega )\pi (d\theta ),\text{ \ \ \ }%
\forall m=1,2,  \label{34} \\
\langle \lambda _{1}\lambda _{2}\rangle _{_{\mathcal{I}}}^{(b_{1},b_{2})}
&=&\int_{F}\int_{\Omega }f_{1}(\theta ,\omega ,b_{1})f_{2}(\theta ,\omega
,b_{2})\nu _{a}^{(b_{1},b_{2})}(d\omega )\pi (d\theta )  \notag
\end{eqnarray}%
in a state $\mathcal{I}=(\Theta ,\mathcal{F}_{\Theta },\pi )$ with the
support on $F\in \mathcal{F}_{\Theta }$ satisfy the original Bell inequality
(\ref{28}) whenever 
\begin{eqnarray}
\int_{\Lambda _{2}}\lambda _{2}(\mathrm{\Pi }_{2,\omega }^{(b_{1})}(d\lambda
_{2}))(\theta ) &=&\pm \int_{\Lambda _{1}}\lambda _{1}(\mathrm{\Pi }%
_{1,\omega }^{(b_{1})}(d\lambda _{1}))(\theta )\text{ \ }\Longleftrightarrow 
\text{ \ \ \ }  \label{35} \\
f_{2}(\theta ,\omega ,b_{1}) &=&\pm f_{1}(\theta ,\omega ,b_{1}),  \notag
\end{eqnarray}%
$\pi \times \nu _{a}^{(b_{1},b_{2})}$-almost everywhere on $F\times \Omega .$

\begin{remark}[On perfect correlations/anticorrelations]
It has been generally accepted to consider that the Bell inequality (\ref{28}%
) holds whenever (see Refs. 1, 2) 
\begin{equation}
\langle \lambda _{1}\lambda _{2}\rangle _{_{\mathcal{I}}}^{(b_{1},b_{1})}=%
\pm 1.  \label{36}
\end{equation}%
This Bell's correlation restriction implies that outcomes $\lambda _{1},$ $%
\lambda _{2}$ admit only two values $\pm 1$ and are either perfectly
correlated (plus sign) or anticorrelated (minus sign).\newline
In contrast to Bell's sufficient condition (\ref{36}), the sufficient
condition (\ref{35}) does not impose any restriction on a type of observed
outcomes. Moreover, even in case of $(\pm 1)$-valued outcomes, our
\textquotedblright average\textquotedblright\ condition (\ref{35}) on
marginal generalized observables is more general than the Bell restriction (%
\ref{36}) on the observed outcomes.\newline
Namely, since $\left\vert f_{i}(\theta ,\omega ,b_{1})\right\vert \leq 1,$ $%
\forall i=1,2$, Eqs. (\ref{34}), (\ref{36}) imply 
\begin{equation}
f_{1}(\theta ,\omega ,b_{1})f_{2}(\theta ,\omega ,b_{1})=\pm 1\text{ \ \ \ }%
\Rightarrow \text{ \ \ }f_{1}(\theta ,\omega ,b_{1})=\pm f_{2}(\theta
,\omega ,b_{1}),  \label{37}
\end{equation}%
$\pi \times \nu _{a}^{(b_{1},b_{2})}$-almost everywhere on $F\times \Omega .$
Thus, the validity of Bell's correlation restriction (\ref{36}) implies the
validity of the condition (\ref{35}). \newline
However, the converse of this statement is not true and, for factorizable
generalized observables (\ref{33}), satisfying the condition (\ref{35}), the
correlation function 
\begin{equation}
\langle \lambda _{1}\lambda _{2}\rangle _{_{\mathcal{I}}}^{(b_{1},b_{1})}=%
\pm \int_{F}\int_{\Omega }f_{1}^{2}(\theta ,\omega ,b_{1})\nu
_{a}^{(b_{1},b_{1})}(d\omega )\pi (d\theta )  \label{38}
\end{equation}%
may, in general, take\footnote{%
Recall that, in general, $f_{1}(\theta ,\omega ,b_{1})\ $may admit any value
in $[-1,1]$.} any value in $[0,1]$ - in case of plus sign and any value in $%
[-1,0]$ - in case of minus sign.
\end{remark}

\subsection{Sufficient Condition}

\bigskip

In view of our results in Sec. 3.1, let us now specify a general condition
sufficient for the validity of CHSH-form inequalities under joint
measurements upon a system of any type.

\begin{condition}[Sufficient]
If four joint measurements are represented on a system information space by
the factorizable generalized observables (\ref{31}), then the corresponding
expectation values satisfy a CHSH-form inequality for any system initial
information state with the support on $F\in \mathcal{F}_{\Theta }.$
\end{condition}

\begin{remark}[On the existence of formal LHV models]
It is necessary to underline that since, in general, a probability
distribution $\nu _{a_{1},a_{2}}^{(b_{1},b_{2})}$ in Eq. (\ref{31}) depends
on settings of joint measurements, the representations (\ref{32}) for the
expectation values in a state $\mathcal{I}$ hold only under these joint
measurements. \newline
Therefore, the validity of the representations (\ref{32}) for an information
state $\mathcal{I}$ does not, in general, mean the existence for this state
of a local hidden variable (LHV) model.\newline
If, however, a probability distribution $\nu $ does not depend on
measurement settings and the representations (\ref{32}) hold for any
measurement parameters then a state $\mathcal{I}$ admits an LHV model (in
general, formal\footnote{%
Recall that, in a Bell \textit{LHV} model (see Refs. 1 - 3), the values of
measurable functions represent the observed outcomes and, therefore, this
model is classical.}). The existence of formal \textit{LHV} models for some
bipartite quantum states was first specified in Ref. 11. \newline
Notice that \textit{the existence of a formal LHV model under joint
measurements does not imply \textquotedblright
classicality\textquotedblright\ of the observed system.}
\end{remark}

\begin{remark}[On possible reasons for the violation ]
Under joint measurements on an information state $\mathcal{I}$, the
violation of a CHSH-form inequality may happen if: (i) generalized
observables, describing these joint measurements, do not have the
factorizable form (\ref{31}); (ii) generalized observables, describing these
joint measurements, are factorizable on a set $F\in \mathcal{F}_{\Theta }$
while the support of a state $\mathcal{I}$ is out of $F.$\newline
The latter is just a general situation under Alice/Bob joint measurements on
a bipartite quantum system and we discuss this in Sec. 4.1.
\end{remark}

\begin{remark}[On ''local realism'']
As we show in Sec. 4, Condition 1 does not, in general, represent
mathematically the physical concept of ''local realism''\ under joint
measurements. Therefore, the violation of a CHSH-form inequality cannot be
linked with the violation of ''local realism''.\bigskip
\end{remark}

\bigskip

\bigskip

\section{GENERAL \textquotedblright LOCAL REALISM\textquotedblright\
CONDITION}

\bigskip

In a general setting, consider now a joint experiment, with outcomes in $%
\Lambda _{1}\times \Lambda _{2}$, performed on a system of any type
represented initially by an information space $(\Theta ,\mathcal{F}_{\Theta
}).$

Let a set-up of the marginal experiment with outcomes in $\Lambda _{1}$ be
characterized by a parameter "$a$" while a set-up of the marginal experiment
with outcomes in $\Lambda _{2}$ - by a parameter "$b$". In this setting, the
set-up of a joint experiment is specified by a pair $(a,b)$ and we further
denote by $\Pi ^{(a,b)}$ a generalized observable, with an outcome set $%
\Lambda _{1}\times \Lambda _{2},$ representing this joint experiment on $%
(\Theta ,\mathcal{F}_{\Theta })$, and by $\mu ^{(a,b)}(\cdot ;\mathcal{I)}$
- the probability distribution of outcomes if a system is initially in an
information state $\mathcal{I}.$ The marginal probability distributions 
\begin{eqnarray}
\mu _{1}^{(a,b)}(B_{1};\mathcal{I}) &:&=\mu ^{(a,b)}(B_{1}\times \Lambda
_{2};\mathcal{I}),\text{ \ \ }\forall B_{1}\subseteq \Lambda _{1},
\label{39} \\
\mu _{2}^{(a,b)}(B_{2};\mathcal{I}) &:&=\mu ^{(a,b)}(\Lambda _{1}\times
B_{2};\mathcal{I}),\text{ \ \ }\forall B_{2}\subseteq \Lambda _{2},  \notag
\end{eqnarray}
describe the statistics of outcomes under the marginal experiments with
outcomes in $\Lambda _{1}$ and $\Lambda _{2},$ respectively. Recall that the
marginal experiments are represented on $(\Theta ,\mathcal{F}_{\Theta })$ by
the marginal generalized observables $\Pi _{1}^{(a,b)}$ and $\Pi
_{2}^{(a,b)} $ (see Eq. (\ref{3})).

If, under the specified joint experiment, the physical concept of
\textquotedblright \emph{local realism}\textquotedblright \emph{\ }(see Ref.
5, page 160) is not violated then, for any information state $\mathcal{I},$
the marginal probability distribution $\mu _{1}^{(a,b)}(\cdot ;\mathcal{I)}$
must not depend on a parameter $b$ while the marginal probability
distribution $\mu _{2}^{(a,b)}(\cdot ;\mathcal{I})$ must not depend on a
parameter $a,$ that is: 
\begin{equation}
\mu _{1}^{(a,b)}(\cdot ;\mathcal{I)}=\mu _{1}^{(a)}(\cdot ;\mathcal{I)},%
\text{ \ \ \ }\mu _{2}^{(a,b)}(\cdot ;\mathcal{I)}=\mu _{2}^{(b)}(\cdot ;%
\mathcal{I)},  \label{40}
\end{equation}%
for any state $\mathcal{I}$.

For short, we further refer to such joint experiments as \emph{%
\textquotedblright locally realistic\textquotedblright . }

Due to Eqs. (\ref{2}) and (\ref{40}), we have the following necessary and
sufficient condition for a joint generalized observable $\Pi ^{(a,b)}$ to
represent a \textquotedblright locally realistic\textquotedblright\ joint
experiment.

\begin{condition}[On \textquotedblright local realism\textquotedblright ]
A joint generalized observable $\Pi ^{(a,b)},$ with an outcome set $\Lambda
_{1}\times \Lambda _{2},$ represents a \textquotedblright locally
realistic\textquotedblright\ joint experiment iff each of its marginal
generalized observables depends only on a set-up of the corresponding
marginal experiment, that is:%
\begin{eqnarray}
\Pi ^{(a,b)}(B_{1}\times \Lambda _{2}) &=&\Pi _{1}^{(a)}(B_{1}),\text{ \ \ \ 
}\forall B_{1}\subseteq \Lambda _{1},  \label{41} \\
\Pi ^{(a,b)}(\Lambda _{1}\times B_{2}) &=&\Pi _{2}^{(b)}(B_{2}),\text{ \ \ \ 
}\forall B_{2}\subseteq \Lambda _{2}.  \notag
\end{eqnarray}
\end{condition}

\smallskip

For short, we call a joint generalized observable satisfying Condition 2 as 
\emph{\textquotedblright locally realistic\textquotedblright }.\smallskip

\begin{example}
Consider a\ generalized observable $\Pi _{1}^{(a)}\times \Pi _{2}^{(b)}$
which is product on all of $\Theta $ and a generalized observable $%
\int_{\Omega }\Pi _{1,\omega }^{(a)}\times \Pi _{2,\omega }^{(b)}\nu
(d\omega )$ which is factorizable on all of $\Theta $. Due to Condition 2,
these joint generalized observables are ''locally realistic''. In
particular, an ''image''\ joint generalized observable (\ref{10}),
describing a classical joint measurement, has a product form and, hence, is
''locally realistic''.
\end{example}

\bigskip

However, in a general case, a \textquotedblright locally
realistic\textquotedblright\ joint generalized observable (\ref{41}) is not
necessarily product or factorizable.\emph{\ }

\emph{Non-factorizable "locally realistic" generalized observables do not
satisfy Condition 1. That is why, under "locally realistic" joint
measurements described by these generalized observables, a CHSH-form
inequality does not need to hold.}

The latter is just a general situation under "locally
realistic\textquotedblright\ joint measurements on a bipartite quantum
system.

\subsection{Quantum joint Measurements}

\bigskip

In the quantum case, a system is described in terms of a separable complex
Hilbert space $\mathcal{K}.$ Denote by $\mathcal{R}_{\mathcal{K}}$ the set
of all density operators $\rho $ on a Hilbert space $\mathcal{K}$.

For a quantum system, we take an information space to be represented by $(%
\mathcal{R}_{\mathcal{K}},\mathcal{B}_{\mathcal{R}_{\mathcal{K}}})$ where $%
\mathcal{B}_{\mathcal{R}_{\mathcal{K}}}$ is the Borel $\sigma $-algebra%
\footnote{%
Representing the trace on $\mathcal{R}_{\mathcal{K}}$ of the Borel $\sigma $%
-algebra on the Banach space of trace class operators on $\mathcal{K}.$} on $%
\mathcal{R}_{\mathcal{K}}$.

Any \emph{quantum} generalized observable on\ $(\mathcal{R}_{\mathcal{K}},%
\mathcal{B}_{\mathcal{R}_{\mathcal{K}}}),$ with an outcome set $\Lambda ,$
is \emph{convex linear} in $\rho $ and is given by (see Ref. 7, section
5.2): 
\begin{equation}
(\Pi _{q}(B))(\rho )=\mathrm{tr}[\rho M(B)],\text{ \ \ }\forall \rho \in 
\mathcal{R}_{\mathcal{K}},\text{ \ \ }\forall B\subseteq \Lambda ,
\label{42}
\end{equation}
where $M$ is a normalized measure with values $M(B),$ $\forall B,$ that are
positive bounded linear operators on $\mathcal{K},$ that is, a positive
operator-valued ($\emph{POV)}$ measure\footnote{%
For the notion of a POV measure, see Refs. 8 - 10.}.

Since any quantum generalized observable is convex linear in $\rho \in 
\mathcal{R}_{\mathcal{K}},$ under a quantum measurement, any two initial
quantum information states $(\mathcal{R}_{\mathcal{K}},\mathcal{B}_{\mathcal{%
R}_{\mathcal{K}}},\pi _{1})$ and $(\mathcal{R}_{\mathcal{K}},\mathcal{B}_{%
\mathcal{R}_{\mathcal{K}}},\pi _{2}),$ satisfying the relation $\int_{%
\mathcal{R}_{\mathcal{K}}}\rho \pi _{1}(d\rho )=\int_{\mathcal{R}_{\mathcal{K%
}}}\rho \pi _{2}(d\rho ),$ give the same information on the statistics of
the observed outcomes.

Consider the description of a joint quantum measurement with outcomes in $%
\Lambda _{1}\times \Lambda _{2}.$

Suppose that, under this joint measurement, a set-up of the marginal
measurement with outcomes in $\Lambda _{1}$ is specified by a parameter $"a"$
while a set-up of a marginal measurement with outcomes in $\Lambda _{2}$ -
by a parameter $"b".$ Let $\Pi _{q}^{(a,b)}$ be a generalized observable,
representing this joint quantum measurement on the quantum information space 
$(\mathcal{R}_{\mathcal{K}},\mathcal{B}_{\mathcal{R}_{\mathcal{K}}})$ and $%
M^{(a,b)}$ be the POV measure uniquely corresponding to $\Pi _{q}^{(a,b)}$
due to (\ref{42}). The marginal POV measures $M^{(a,b)}(B_{1}\times \Lambda
_{2}),$ $\forall B_{1}\subseteq \Lambda _{1},$ and $M^{(a,b)}(\Lambda
_{1}\times B_{2}),$ $\forall B_{2}\subseteq \Lambda _{2},$ describe the
corresponding marginal quantum measurements.

From Condition 2 and Eq. (\ref{42}) it follows that if a quantum joint
measurement is described by a POV measure\emph{\ }$M^{(a,b)}$ satisfying the
relations: 
\begin{equation}
M^{(a,b)}(B_{1}\times \Lambda _{2})=M_{1}^{(a)}(B_{1}),\text{ \ \ \ }%
M^{(a,b)}(\Lambda _{1}\times B_{2})=M_{2}^{(b)}(B_{2}),  \label{43}
\end{equation}%
for any outcome subsets $B_{1}\subseteq \Lambda _{1},B_{2}\subseteq \Lambda
_{2},$ then this joint quantum measurement is \emph{"locally
realistic\textquotedblright . }

Consider an example of a quantum "locally realistic" joint measurement.

\begin{example}[Alice/Bob joint quantum measurement]
Consider a bipartite quantum system described in terms of a separable
complex Hilbert space $\mathcal{H}_{1}\otimes \mathcal{H}_{2}$ and let a\
joint quantum measurement on this system be represented by the POV measure 
\begin{equation}
M_{1}^{(a)}(B_{1})\otimes M_{2}^{(b)}(B_{2}),  \label{44}
\end{equation}%
for any $B_{1}\subseteq \Lambda _{1},$ $B_{2}\subseteq \Lambda _{2}.$ This
POV measure satisfies the condition (\ref{43}) and, hence, represents a
\textquotedblright locally realistic\textquotedblright\ joint quantum
measurement. For convenience, $\Lambda _{1}$ and $\Lambda _{2}$ are referred
to as sets of outcomes on the \textquotedblright sides\textquotedblright\ of
Alice and Bob, respectively. \newline
Thus, any Alice/Bob joint quantum measurement is \textquotedblright locally
realistic\textquotedblright .
\end{example}

\smallskip

Due to Eqs. (\ref{42}), (\ref{44}), under a quantum Alice/Bob joint
measurement, the joint generalized observable has the form: 
\begin{equation}
(\Pi _{q}^{(a,b)}(B_{1}\times B_{2}))(\rho )=\mathrm{tr}[\rho
(M_{1}^{(a)}(B_{1})\otimes M_{2}^{(b)}(B_{2}))],\text{ \ \ }\forall \rho \in 
\mathcal{R}_{\mathcal{H}_{1}\otimes \mathcal{H}_{2}}.  \label{45}
\end{equation}
On any separable density operator $\rho _{S}=$ $\sum_{j}\gamma _{j}\rho
_{1}^{(j)}\otimes \widetilde{\rho }_{2}^{(j)},\ \gamma _{j}>0,\
\sum_{j}\gamma _{j}=1,$ this generalized observable admits a representation 
\begin{equation}
(\Pi _{q}^{(a,b)}(B_{1}\times B_{2}))(\rho _{S})=\sum_{j}\gamma _{j}\mathrm{%
tr}[\rho _{1}^{(j)}M_{1}^{(a)}(B_{1})]\text{ }\mathrm{tr}[\rho
_{2}^{(j)}M_{2}^{(b)}(B_{2})]  \label{46}
\end{equation}
and, hence, is factorizable. Due to Condition 1, under four quantum
Alice/Bob joint measurements 
\begin{equation}
M^{(a_{k},b_{m})}(B_{1}\times B_{2})=M_{1}^{(a_{k})}(B_{1})\otimes
M_{2}^{(b_{m})}(B_{2}),\text{ \ \ \ }\forall k,m=1,2,  \label{47}
\end{equation}
performed on a separable quantum state, a CHSH-form inequality is satisfied.

On any density operator $\rho \in \mathcal{R}_{\mathcal{H}_{1}\otimes 
\mathcal{H}_{2}}$, four joint generalized observables (\ref{45}) do not, in
general, admit the factorizable representations (\ref{31}). That is why, 
\emph{under} \emph{Alice/Bob joint measurements on an arbitrary bipartite
quantum state, Condition 1 is not, in general, fulfilled and a CHSH-form
inequality may be violated - though quantum Alice/Bob joint measurements are
"locally realistic".}

\bigskip

\bigskip

\section{QUANTUM ANALOGS OF THE BELL INEQUALITY}

\bigskip

In this section, for an arbitrary quantum state $\rho $ of two identical
sub-systems, we introduce quantum Bell-form inequalities under Alice/Bob
joint measurements.

In case of identical quantum sub-systems, $\mathcal{H}_{1}=\mathcal{H}_{2}=%
\mathcal{H}$ and a bipartite state $\rho $ on $\mathcal{H}\otimes \mathcal{H}
$ must be symmetric, that is: $\mathrm{S}_{2}\rho =\rho ,$ where $\mathrm{S}%
_{2}$ is the symmetrization operator on the space of bounded linear operators%
\textbf{\ }on $\mathcal{H}\otimes \mathcal{H}$.

Moreover, each of marginal POV measures must have a symmetrized tensor
product form and be specified by a set of outcomes on the \textquotedblright
side\textquotedblright\ of Alice or Bob but not by the \textquotedblright
side\textquotedblright\ of the tensor product. The latter means that, for an
Alice/Bob joint quantum measurement on identical sub-systems, the POV
measure has the form: 
\begin{eqnarray}
M^{(a,b)}(B_{1}\times B_{2}) &=&\{M_{1}^{(a)}(B_{1})\otimes
M_{2}^{(b)}(B_{2})\}_{sym}  \label{48} \\
&:&=\frac{1}{2}\{M_{1}^{(a)}(B_{1})\otimes
M_{2}^{(b)}(B_{2})+M_{2}^{(b)}(B_{2})\otimes M_{1}^{(a)}(B_{1})\},  \notag
\end{eqnarray}
for any outcome subsets $B_{1}\subseteq \Lambda _{1},$ $B_{2}\subseteq
\Lambda _{2}.$

For simplicity, we further suppose that outcomes $|\lambda _{1}|$ $\leq 1$
and $|\lambda _{2}|$ $\leq 1.$

Under an Alice/Bob joint quantum measurement (\ref{48}) on a symmetric state 
$\rho $, the expectation values\ (\ref{7}) are given by\footnote{%
Notice that $\mathrm{tr}[\sigma \{W_{1}\otimes W_{2}\}_{sym}]$ $=\mathrm{tr}%
[\sigma (W_{1}\otimes W_{2})]$, for any symmetric trace class operator $%
\sigma .$}: 
\begin{eqnarray}
\langle \lambda _{1}\lambda _{2}\rangle _{\rho }^{(a,b)} &=&\int_{\Lambda
_{1}\times \Lambda _{2}}\lambda _{1}\lambda _{2}\mathrm{tr}[\rho
\{M_{1}^{(a)}(d\lambda _{1})\otimes M_{2}^{(b)}(d\lambda _{2})\}_{sym}]
\label{49} \\
&=&\mathrm{tr}[\rho (A_{1}^{(a)}\otimes A_{2}^{(b)})],  \notag
\end{eqnarray}
where 
\begin{equation}
A_{1}^{(a)}=\int_{\Lambda _{1}}\lambda _{1}M_{1}^{(a)}(d\lambda _{1}),\text{
\ \ \ }A_{2}^{(b)}=\int_{\Lambda _{2}}\lambda _{2}M_{2}^{(b)}(d\lambda _{2})
\label{50}
\end{equation}
are self-adjoint bounded linear operators on $\mathcal{H},$ with the
operator norms $||A_{1}^{(a)}||$ $\leq 1,$ $||A_{2}^{(b)}||$ $\leq 1\mathrm{.%
}$

For a state $\rho ,$ introduce a representation 
\begin{equation}
\rho =\eta (\tau ,\widetilde{\tau })+\sigma _{\rho }^{(\eta )}  \label{51}
\end{equation}
via a separable density operator 
\begin{equation}
\eta (\tau ,\widetilde{\tau }):=\frac{1}{2}\sum_{j}\gamma _{j}(\tau
_{j}\otimes \widetilde{\tau }_{j}+\widetilde{\tau }_{j}\otimes \tau _{j}),%
\text{ \ \ }\gamma _{j}>0,\ \ \sum_{j}\gamma _{j}=1,  \label{52}
\end{equation}
where $\{\tau _{j}\}_{j=1}^{N}$ and $\{\widetilde{\tau }_{j}\}_{j=1}^{N}$, $%
N\leq \infty ,$ are any families of density operators on $\mathcal{H}.$ In
Eq. (\ref{51}), the operator $\sigma _{\rho }^{(\eta )}$ is symmetric,
self-adjoint, trace class and its trace norm $||\sigma _{\rho }^{(\eta
)}||_{1}$ characterizes a "distance" between $\rho $ and a separable state $%
\eta (\tau ,\widetilde{\tau }).$

For concreteness, we further refer to Eq. (\ref{51}) as a $(\tau ,\widetilde{%
\tau })$-representation of $\rho .$

Substituting Eq. (\ref{51}) into Eq. (\ref{49}) and using the inequality (%
\ref{15}), we derive that, for a state $\rho ,$ the relation 
\begin{eqnarray}
&&\left\vert \langle \lambda _{1}\lambda _{2}\rangle _{\rho
}^{(a,b_{1})}-\langle \lambda _{1}\lambda _{2}\rangle _{\rho
}^{(a,b_{2})}-\langle z\rangle _{\sigma _{\rho }^{(\eta )}}\right\vert
\label{53} \\
&\leq &1-\frac{1}{2}\sum_{j}\gamma _{j}\{\mathrm{tr}[\tau
_{j}A_{2}^{(b_{1})}]\text{\textrm{tr}}[\tau _{j}A_{2}^{(b_{2})}]+\mathrm{tr}[%
\widetilde{\tau }_{j}A_{2}^{(b_{1})}]\text{\textrm{tr}}[\widetilde{\tau }%
_{j}A_{2}^{(b_{2})}]\}  \notag
\end{eqnarray}
holds for any $(\tau ,\widetilde{\tau })$-representation of $\rho .$ Here%
\footnote{%
We use the bound $|tr[\sigma W]|\leq ||\sigma ||_{1}||W||,$ valid for any
trace class operator $\sigma $ and any bounded linear operator $W.$}, 
\begin{equation}
\langle z\rangle _{\sigma _{\rho }^{(\eta )}}:=\mathrm{tr}[\sigma _{\rho
}^{(\eta )}(A_{1}^{(a)}\otimes (A_{2}^{(b_{1})}-A_{2}^{(b_{2})}))],\text{ \
\ \ }|\text{ }\langle z\rangle _{\sigma _{\rho }^{(\eta )}}|\text{ }\leq
||\sigma _{\rho }^{(\eta )}||_{1}||A_{2}^{(b_{1})}-A_{2}^{(b_{2})}||.
\label{54}
\end{equation}

\begin{proposition}[Quantum analogs]
Let, under Alice/Bob joint measurements (\ref{48}) with outcomes $|\lambda
_{1}|$ $\leq 1$ and $|\lambda _{2}|$ $\leq 1$, the marginal POV measures
satisfy the condition 
\begin{equation}
\int_{\Lambda _{2}}\lambda _{2}M_{2}^{(b_{1})}(d\lambda _{2})=\int_{\Lambda
_{1}}\lambda _{1}M_{1}^{(b_{1})}(d\lambda _{1})\text{ \ \ }%
\Longleftrightarrow \text{ \ \ }A_{2}^{(b_{1})}=A_{1}^{(b_{1})}.  \label{55}
\end{equation}
Then a quantum state $\rho $ on $\mathcal{H}\otimes \mathcal{H}$ satisfies
the inequality 
\begin{equation}
\left\vert \langle \lambda _{1}\lambda _{2}\rangle _{\rho
}^{(a,b_{1})}-\langle \lambda _{1}\lambda _{2}\rangle _{\rho
}^{(a,b_{2})}\right\vert \leq \gamma _{\rho }^{(\eta )}-\langle \lambda
_{1}\lambda _{2}\rangle _{\widetilde{\eta }(\tau ,\widetilde{\tau }%
)}^{(b_{1},b_{2})},  \label{56}
\end{equation}
for every $(\tau ,\widetilde{\tau })$-representation (\ref{51}) of $\rho $.
Here, 
\begin{eqnarray}
\gamma _{\rho }^{(\eta )} &=&1+||\rho -\eta (\tau ,\widetilde{\tau }%
)||_{1}||A_{2}^{(b_{1})}-A_{2}^{(b_{2})}||  \label{57} \\
&\leq &1+2||\rho -\eta (\tau ,\widetilde{\tau })||_{1},  \notag \\
\widetilde{\eta }(\tau ,\widetilde{\tau }) &=&\frac{1}{2}\sum_{j}\gamma
_{j}(\tau _{j}\otimes \tau _{j}+\widetilde{\tau }_{j}\otimes \widetilde{\tau 
}_{j}).  \notag
\end{eqnarray}
\end{proposition}

\begin{proof}
We use the inequality (\ref{53}), Eq. (\ref{54}), the condition (\ref{55})
and then the notation (\ref{57}).
\end{proof}

\smallskip

The inequality (\ref{56}) describes the relation between the expectation
values under three Alice/Bob joint quantum measurements and we refer to it
as a quantum analog of the Bell inequality.

Let us now specify the inequality (\ref{56}) in case of a separable
bipartite state. For a (symmetric) separable quantum state $\rho _{S},$
there always exists a representation 
\begin{equation}
\rho _{S}=\frac{1}{2}\sum_{j}\gamma _{j}\{\tau _{j}\otimes \widetilde{\tau }%
_{j}+\widetilde{\tau }_{j}\otimes \tau _{j}\}+\sigma _{\rho _{S}}^{(s)},%
\text{ \ \ }\gamma _{j}>0,\text{ \ \ \ }\sum_{i}\gamma _{j}=1,  \label{58}
\end{equation}%
where $||\sigma _{\rho _{S}}^{(s)}||_{1}=0.$ Hence, for a separable state $%
\rho _{S},$ the inequality (\ref{56}), corresponding to the representation (%
\ref{58}), takes the form: 
\begin{eqnarray}
\left\vert \langle \lambda _{1}\lambda _{2}\rangle _{\rho
_{S}}^{(a,b_{1})}-\langle \lambda _{1}\lambda _{2}\rangle _{\rho
_{S}}^{(a,b_{2})}\right\vert &\leq &1-\langle \lambda _{1}\lambda
_{2}\rangle _{\widetilde{\rho }_{S}}^{(b_{1},b_{2})},  \label{59} \\
\widetilde{\rho }_{S} &=&\frac{1}{2}\sum_{j}\gamma _{j}(\tau _{j}\otimes
\tau _{j}+\widetilde{\tau }_{j}\otimes \widetilde{\tau }_{j}),  \notag
\end{eqnarray}%
and coincides with the inequality (40) introduced in Ref. 12.

Suppose further that a separable quantum state $\rho _{S}$ is of the special
form: 
\begin{equation}
\rho _{S}=\sum_{j}\gamma _{j}\tau _{j}\otimes \tau _{j}+\sigma _{\rho
_{S}}^{(s)},\text{ \ \ \ \ }||\sigma _{\rho _{S}}^{(s)}||_{1}=0.  \label{60}
\end{equation}
For this state $\widetilde{\rho }_{S}=\rho _{S}-\sigma _{\rho _{S}}^{(s)}$
and $\langle \lambda _{1}\lambda _{2}\rangle _{\widetilde{\rho }%
_{S}}^{(b_{1},b_{2})}=\langle \lambda _{1}\lambda _{2}\rangle _{\rho
_{S}}^{(b_{1},b_{2})}.$

\begin{corollary}
Under Alice/Bob joint measurements (\ref{48}), satisfying the condition (\ref%
{55}), the perfect correlation form of the original Bell inequality 
\begin{equation}
\left\vert \langle \lambda _{1}\lambda _{2}\rangle _{\rho
_{S}}^{(a,b_{1})}-\langle \lambda _{1}\lambda _{2}\rangle _{\rho
_{S}}^{(a,b_{2})}\right\vert \leq 1-\langle \lambda _{1}\lambda _{2}\rangle
_{\rho _{S}}^{(b_{1},b_{2})}  \label{61}
\end{equation}
holds for any separable quantum state of the special form (\ref{60}).
\end{corollary}

It is necessary to underline that the operator condition (\ref{55}) on
marginal POV measures is always true under Alice and Bob projective
measurements of the same quantum observable on both sides. That is why, a
separable state of the special form (\ref{60}) satisfies the perfect
correlation form (\ref{61}) of the original Bell inequality under any three
projective quantum measurements of Alice and Bob, specified on Alice and Bob
sides by pairs of bounded quantum observables: $(A^{(a)},A^{(b_{1})}),$ $%
(A^{(a)},A^{(b_{2})})$ and $(A^{(b_{1})},A^{(b_{2})})$. Notice\footnote{%
See also Remark 2.} that, satisfying the perfect correlation form of the
Bell inequality for any bounded quantum observables $A^{(a)},$ $A^{(b_{1})},$
$A^{(b_{2})}$, a bipartite quantum state (\ref{60}) does not necessarily
exhibit perfect correlations.

\bigskip

\bigskip

\section{ON LOCALITY\ OF QUANTUM\ INTERACTIONS}

\bigskip

In the present paper, we discuss in a very general setting the description
of joint experiments performed on a system of any type.

Mathematically, any joint experiment is described by the notion of a joint
generalized observable and this notion does not include any specifications
on whether or not marginal experiments are separated in space and in time.

The main results of our paper indicate:

\begin{itemize}
\item The physical concept of ''local realism''\ can be expressed in
mathematical terms for a joint experiment upon a system of any type. The
generally accepted mathematical specification of this concept in the frame
of a hidden variable model corresponds only to a particular case of joint
experiments represented by factorizable generalized observables;

\item The general sufficient condition for a CHSH-form inequality to hold is
not equivalent to the condition on \textquotedblright local
realism\textquotedblright\ under joint experiments. Therefore, the violation
of a CHSH-form inequality in the quantum case does not point to the
violation of the physical concept of \textquotedblright local
realism\textquotedblright ;

\item Quantum joint experiments of the Alice/Bob type are \textquotedblright
locally realistic\textquotedblright . However, under these
\textquotedblright locally realistic\textquotedblright\ joint experiments,
the sufficient condition for a CHSH-form inequality to hold is not satisfied
for any bipartite quantum state;

\item Quantum analogs of the original Bell inequality, derived in this
paper, specify the relation between the statistical data observed under
quantum \textquotedblright locally realistic\textquotedblright\ joint
experiments on an arbitrary bipartite quantum state.
\end{itemize}

\emph{In the light of these results, we argue that the violation of
Bell-type inequalities in the quantum case cannot be a valid argument in the
discussion on locality or non-locality of quantum interactions.}

\emph{\medskip }

\noindent \textbf{Acknowledgments. }I am grateful to Klaus Molmer and Asher
Peres for useful discussions. This work is partially supported by MaPhySto -
A Network in Mathematical Physics and Stochastics, funded by The Danish
National Research Foundation.

\end{document}